\documentclass[onecolumn,tighten,apj]{emulateapj}

\usepackage{amssymb,amsmath,times,graphicx}

\newcommand{\txa}{{\text{a}}}
\newcommand{\txd}{{\text{d}}}
\renewcommand{\leq}{\leqslant}
\renewcommand{\geq}{\geqslant}
\newcommand{\revision}[1]{#1}

\begin{document}

\title{On the universality of the global slope -- anisotropy inequality}

\author{Emmanuel Van Hese, Maarten Baes and Herwig Dejonghe}
\affil{Sterrenkundig Observatorium, Universiteit Gent, Krijgslaan 281 S9, B-9000 Gent, Belgium}
\email{emmanuel.vanhese@gmail.com}
\email{maarten.baes@ugent.be}
\email{herwig.dejonghe@ugent.be}



\submitted{Draft version \today; Accepted for publication in ApJ}
\label{firstpage}

\begin{abstract}
  Recently, some intriguing results have lead to speculations whether
  the central density slope -- velocity dispersion anisotropy
  inequality (An \& Evans) actually holds at all radii for spherical
  dynamical systems. We extend these studies by providing a complete
  analysis of the global slope -- anisotropy inequality for all
  spherical systems in which the augmented density is a separable
  function of radius and potential. We prove that these systems indeed
  satisfy the global inequality if their central anisotropy is
  $\beta_0\leq 1/2$. Furthermore, we present several systems with
  $\beta_0 > 1/2$ for which the inequality does not hold, thus
  demonstrating that the global density slope -- anisotropy inequality
  is not a universal property. This analysis is a significant step
  towards an understanding of the relation for general spherical
  systems.
\end{abstract}


\maketitle

\section{Introduction}
\label{introduction.sec}
\defcitealias{2007A&A...471..419B}{Paper~I}
\defcitealias{2009ApJ...690.1280V}{Paper~II} 

Theoretical dynamical models continue to play a key role in stellar
dynamics, as understanding their underlying structure helps shed light
on the properties of numerical and observational stellar systems and
dark matter haloes. In this paper, we focus our attention on the
relation between the density slope $\gamma(r)$ and the velocity
anisotropy profile $\beta(r)$, which has attracted renewed interest
lately. \revision{As is well known, \cite{2006AJ....131..782A} proved
  that the central inequality $\gamma_0\geqslant 2\beta_0$ is a
  necessary condition for the positivity of the distribution function
  (DF) of a spherical system. More recently however
  \cite{2010MNRAS.401.1091C,2010MNRAS.408.1070C} showed that
  $\gamma(r)\geqslant 2\beta(r)$ at all radii (hereafter called the
  Global Density Slope -- Anisotropy Relation, GDSAI) is a necessary
  condition for positivity of the DF, if $\beta_0\leqslant 1/2$, in
  the families of multi-component Osipkov-Merritt
  \citep{1979PAZh....5...77O,1985AJ.....90.1027M}, Cuddeford 
  \citep{1991MNRAS.253..414C}, and Cuddeford-Louis models
  \citep{1995MNRAS.275.1017C}, as well as for the Plummer models of
  \citet{1987MNRAS.224...13D}, the Hernquist models of
  \citet{2002A&A...393..485B}, and the models we introduced in
  \cite{2007A&A...471..419B} (hereafter
  \citetalias{2007A&A...471..419B}).  Their proof is based on the fact
  that all these models are characterized by having a separable
  augmented density (see Section~\ref{augdens.sec}). They also note
  that currently, no counter-examples of the GDSAI are known, but
  remark that in the case of Cuddeford models with a central
  anisotropy $\beta_0 > 1/2$ the GDSAI is only a sufficient condition,
  so that possible counter-examples could be found in this range of
  values.}

These results pose the question under which conditions the GDSAI holds
for all spherical systems. \revision{In this paper, we make important
  advancements by providing a complete analysis of the GDSAI for all
  well-behaved systems with a separable augmented density.}  This
group includes all aforementioned models, as well as the hypervirial
models of \citet{2005MNRAS.360..492E}, the $\gamma$-models of
\citet{2007MNRAS.375..773B} and the Dehnen-McLaughlin systems
discussed in \cite{2009ApJ...690.1280V} (hereafter
\citetalias{2009ApJ...690.1280V}), among others. \revision{First, we show that
the GDSAI holds for all separable systems, if $\beta_0\leqslant 1/2$,
by proving an equivalent criterion formulated by
\cite{2010MNRAS.408.1070C}. In this manner, we extend
  their previous results}. Our analysis also
reveals some very peculiar properties of separable systems.
Furthermore, we show that counter-examples of the GDSAI do exist for
separable systems with $\beta_0 > 1/2$, \revision{in other words, we
  demonstrate that the GDSAI is not a universal property}. However,
the velocity distributions of these models are extreme, and all
counter-examples are very likely dynamically unstable.

First, we outline in Section~\ref{models.sec} the general concepts of
spherical dynamical models.  In Section~\ref{augdens.sec}, we describe
the augmented density framework. In Section~\ref{analysis.sec}, we
give our analysis of the GDSAI for separable systems: we prove the
inequality for models with $\beta_0\leqslant 1/2$, and we present
three counter-examples with $\beta_0 > 1/2$. Finally, we discuss our
results in Section~\ref{conclusions.sec}.

\section{Spherical dynamical models}
\label{models.sec}
The dynamical structure of a spherical gravitational equilibrium
system, governed by a positive potential $\psi(r)$, is completely
determined by the non-negative phase-space distribution function (DF)
$F(\vec{r},\vec{v})$. For spherical systems, this DF is a function
$F(E,L)$ of the isolating integrals, the binding energy $E$ and the
angular momentum $L$:
\begin{align}
E &= \psi(r) - \frac{1}{2}v_r^2 - \frac{1}{2}v_T^2,\\
L &= r\,v_T,
\end{align}
with 
\begin{equation}
v_T = \sqrt{v_\theta^2 + v_\varphi^2},
\end{equation}
the transverse velocity. From the DF, the velocity moments
\begin{equation}
  \mu_{2n,2m}(r) = 2\pi M \iint
  F(E,L)\,v_r^{2n}\,v_T^{2m+1}\,\txd v_r\,\txd v_T.
\label{momdf}
\end{equation}
can be obtained, with $M$ the total mass of the system. In particular,
the density and the second-order moments are
\begin{equation}
  \rho(r) = \mu_{00}(r),\qquad \rho\sigma_r^2(r) = \mu_{20}(r), 
  \qquad \rho\sigma_T^2(r) = \mu_{02}(r),
\end{equation}
and $\sigma_T^2(r)=2\sigma_\theta^2(r)$. The density slope and the
velocity anisotropy profile are defined as
\begin{align}
  \gamma(r) &= -\frac{\txd\ln\rho}{\txd \ln r}(r),\\
  \beta(r) &= 1 - \frac{\sigma_\theta^2(r)}{\sigma_r^2(r)}.
\end{align}
Spherical dynamical models satisfy the Jeans equation
\begin{equation}
  \frac{\txd\rho\sigma_r^2}{\txd r}(r) 
  + \frac{2\beta(r)}{r}\rho\sigma_r^2(r)
  =
  \rho(r)\frac{\txd\psi}{\txd r}(r),
\end{equation}
which can be written as
\begin{equation}
\sigma_r^2(r)\left(\gamma(r) - 2\beta(r) + \kappa(r)\right) = v_c^2(r),
\end{equation}
with 
\begin{equation}
  \kappa(r) = -\frac{\txd\ln\sigma_r^2}{\txd \ln r}(r),\qquad v_c^2(r) = -r\frac{\txd\psi}{\txd r}(r).
\end{equation}
Evidently, it follows that
\begin{equation}
\gamma(r) - 2\beta(r) + \kappa(r)\geqslant 0,\qquad \forall\,r.
\end{equation}
\cite{2010MNRAS.401.1091C,2010MNRAS.408.1070C} showed that several
systems (see Introduction) satisfy a stronger condition, the GDSAI
\begin{equation}
\gamma(r) - 2\beta(r) \geqslant 0,\qquad \forall\,r,
\end{equation}
and they pose the question whether this condition holds for all
spherical systems. Naturally, the inequality is valid outside the
radius $r_2$ where $\gamma(r_2)=2$. It is also valid at $r=0$, as was
proven by \cite{2006AJ....131..782A}.  In this paper, we will
investigate the GDSAI for a particular class of systems, namely those
with a separable augmented density.

\section{The augmented density concept}
\label{augdens.sec}
A spherical dynamical system can also described by an augmented
velocity moment (\cite{1986PhR...133..217D};
\citetalias{2007A&A...471..419B}), which extends the moment to an
explicit function $\tilde\mu_{2n,2m}(\psi,r)$ of both the radius and
the gravitational potential. An augmented moment is equivalent to the
DF: the knowledge of one augmented moment determines the entire
system. In particular, we will consider the augmented density
$\tilde\rho(\psi,r)$, and its relationship with the DF is given by
\begin{equation}
  \tilde\rho(\psi,r) 
  = 2\pi M \int_0^\psi \txd E 
  \int_0^{2(\psi-E)} \frac{F(E,rv_T)}{\sqrt{2(\psi-E)-v_T^2}}\,\txd v_T^2.
\label{rhodf}
\end{equation}
This integral equation can in principle be inverted to obtain the DF
by using Laplace-Mellin transforms, although in practice the inversion
is only numerically stable for sufficiently smooth systems. The
strength of the augmented density framework lies in its direct
connection to observable quantities like the velocity moments. For instance,
the augmented velocity dispersion profiles are given by
\begin{align}
    \tilde\sigma_r^2(\psi,r)
    &=
    \frac{1}{\tilde\rho(\psi,r)}
    \int_0^\psi \tilde\rho(\psi',r)\,\txd\psi',
    \label{gensigr:def}
    \\
    \tilde\sigma_T^2(\psi,r)
    &=
    \frac{2}{\tilde\rho(\psi,r)}
    \int_0^\psi
    D_{r^2}
    \left[r^2\,\tilde\rho(\psi',r)\right]
    \txd\psi',
\label{gensigt:def}
\end{align}
where $D_{r^2}$ denotes the derivative with respect to $r^2$. The
observed density and dispersions are then simply recovered from
\begin{align}
  \rho(r) &= \tilde\rho(\psi(r),r),\\
  \sigma_r^2(r) &= \tilde\sigma_r^2(\psi(r),r),\\
  \sigma_T^2(r) &= \tilde\sigma_T^2(\psi(r),r),
\end{align}
and the density slope is
\begin{equation}
\gamma(r) = -\frac{r}{\rho}\frac{\partial\tilde\rho}{\partial r}(\psi(r),r)
- \frac{r}{\rho}\frac{\txd\psi}{\txd r}(r)\,\frac{\partial\tilde\rho}{\partial \psi}(\psi(r),r).
\label{slopegen}
\end{equation}
As remarked in the Introduction, Ciotti \& Morganti have examined the GDSAI in
several systems with a separable augmented density, i.e.\ systems of the form
\begin{equation}
    \tilde\rho(\psi,r)
    =
    f(\psi)\,g(r),\qquad 0 \leqslant \psi \leqslant \psi_0,
\label{splitaugdens}
\end{equation}
with $\psi_0=\psi(0)$. For such models, the dispersion profiles read
\begin{align}
    \tilde\sigma_r^2(\psi)
    &=
    \frac{1}{f(\psi)}
    \int_0^\psi f(\psi')\,\txd\psi',
    \label{sigmarfg}
    \\
    \tilde\sigma_T^2(\psi,r)
    &=
    \left(1+\frac{1}{2}\frac{\txd\ln g}{\txd\ln r}\right)
    \frac{2}{f(\psi)}
    \int_0^\psi f(\psi')\,\txd\psi'.
\end{align}
Note that the radial velocity dispersion is now only a function of $\psi$.
The velocity anisotropy profile of these systems has the simple form
\begin{equation}
    \beta(r)
    =
    -\frac{1}{2}\frac{\txd\ln g}{\txd\ln r}(r).
\label{betadv}
\end{equation}
As we demonstrated in \citetalias{2007A&A...471..419B} and
\citetalias{2009ApJ...690.1280V}, this property provides a very
elegant way to construct dynamical models with a given potential,
density and velocity anisotropy. \revision{Indeed, separable systems
  are completely determined by $\psi(r)$, $\rho(r)$ and $\beta(r)$,
  since $g(r)$ is defined by Eq.~(\ref{betadv}) and, by inverting $\psi(r)$,
  the function $f(\psi)$ follows from
\begin{equation}
  f(\psi) = \frac{\rho(r(\psi))}{g(r(\psi))}.
\end{equation}
However, one still needs to verify whether the corresponding DF is
non-negative everywhere.}  Eq.~(\ref{slopegen}) now reduces to
\begin{equation}
\gamma(r) = -\frac{\txd\ln g}{\txd\ln r}(r) - \frac{\txd\ln \psi}{\txd\ln r}(r)\,
\frac{\txd\ln f}{\txd\ln \psi}(\psi(r)),
\end{equation}
so that we obtain
\begin{equation}
\frac{\txd f}{\txd\psi}(\psi(r)) = 
\frac{f(\psi(r))}{v_c^2(r)} \left( \gamma(r) - 2\beta(r)\right).
\label{gdsai}
\end{equation}
In other words, as remarked by Ciotti \& Morganti, the GDSAI
\begin{equation}
\gamma(r)\geqslant 2\beta(r),\qquad\forall r \geqslant 0, 
\end{equation}
is for separable systems equivalent to the statement
\begin{equation}
\frac{\txd f}{\txd \psi} \geqslant 0,\qquad\forall \ 0 \leqslant \psi \leqslant \psi_0. 
\label{dfdpsi}
\end{equation}
The question thus becomes whether this inequality is valid for all separable systems. In the
following section, we will prove that this is indeed the case, if $\beta_0\leqslant 1/2$.

\section{Analysis of the GDSAI for separable systems}
\label{analysis.sec}

Following the reasoning of \citet{2006AJ....131..782A}, we
\revision{assume} that any well-behaved DF can be written in the form
\begin{equation}
F(E,L) = L^{-2\beta_0}\left(F_0(E) + F_1(E,L)\right),
\end{equation}
with
\begin{equation}
F_1(E,0) \equiv 0, \quad \forall \ 0 \leqslant E \leqslant \psi_0.
\label{dfcond}
\end{equation}
\revision{The function $L^{-2\beta_0}F_0(E)$ in this Ansatz can be
  understood as the leading term of a Laurent series expansion in $L$
  at $L=0$. Towards the center $r \rightarrow 0$, the DF is dominated
  by this term, which has the form of a system with constant
  anisotropy. Consequently, the central anisotropy of the entire model
  indeed corresponds with $\beta_0$.}  Since the DF has to be
non-negative everywhere, it follows immediately that $F_0(E)\geqslant
0 \ \forall E$ is a necessary condition to obtain a physically
meaningful DF.

If we consider separable systems, the corresponding augmented density
then has the form
\begin{equation}
\label{rhocond} 
 \tilde\rho(\psi,r) = f(\psi)\,r^{-2\beta_0}\left(1+g_1(r)\right), 
  \quad\text{with\ \ } g_1(0) = 0.
\end{equation}
Using $u^2 = \frac{v_T^2}{2(\psi-E)}$, the relation between the augmented
density and the DF~(\ref{rhodf}) can be written as
\begin{equation}
  \tilde\rho(\psi,r) = 2\pi\,2^{1/2-\beta_0} r^{-2\beta_0} M\int_0^1 \frac{u^{-2\beta_0}}{\sqrt{1-u^2}}\,\txd u^2
  \int_0^\psi (\psi-E)^{1/2-\beta_0}
  \left(F_0(E) + F_1\left(E,ru\sqrt{2(\psi-E)}\right)\right)\txd E.
 \label{rhodfappend}
\end{equation}
In separable systems, it follows that
\begin{equation}
  f(\psi) = \frac{\tilde\rho(\psi,r)}{g(r)}.
\end{equation}
Since the left-hand side of this equation is independent of the radius
$r$, the right-hand side does not depend on $r$ either. The equality is
therefore valid for all values $r$; in particular, we can take the
limit of $r$ towards the center,
\begin{equation}
  \label{fpsilim}
  f(\psi) = \lim_{r\rightarrow 0}\frac{\tilde\rho(\psi,r)}{g(r)} 
  = \lim_{r\rightarrow 0} r^{2\beta_0}\tilde\rho(\psi,r).
\end{equation}
This property is the key element to prove the GDSAI when $\beta_0 \leq 1/2$:
using (\ref{dfcond}) and (\ref{fpsilim}), it follows from Eq.~(\ref{rhodfappend}) that
\begin{equation}
  f(\psi) = (2\pi)^{3/2}2^{-\beta_0}M \frac{\Gamma(1-\beta_0)}{\Gamma(3/2-\beta_0)}
  \int_0^\psi (\psi-E)^{1/2-\beta_0}\,F_0(E)\,\txd E.
  \label{fpsi}
\end{equation}
\revision{Remarkably, the function $f(\psi)$ thus only depends on
  $F_0(E)$ and $\beta_0$. In other words, for separable systems the
  function $F_1(E,L)$ has no influence on the GDSAI. Concrete examples
  of this behavior are furnished in the systems considered by Ciotti
  \& Morganti. For instance, the equivalent function $B(\psi_\text{T})$ in
  \cite{2010MNRAS.408.1070C} for generalized Cuddeford systems does
  not depend on the anisotropy radius $r_\txa$ (see their Eq.~(13)).}

The value of $\beta_0$ splits our further analysis into three cases:
$\beta_0 < 1/2$, $\beta_0 = 1/2$, and $\beta_0 > 1/2$.

\subsection{Proof for $\beta_0 < 1/2$}
\label{proof.sec}

If $\beta_0 < 1/2$, the derivative of  $f(\psi)$ becomes
\begin{align}
  \frac{\txd f}{\txd \psi}(\psi) &= (2\pi)^{3/2}2^{-\beta_0}M
    \frac{\Gamma(1-\beta_0)}{\Gamma(3/2-\beta_0)}
  \left[
    \lim_{E\rightarrow\psi}(\psi-E)^{1/2-\beta_0}\,F_0(E)\right.\nonumber\\
    \ 
    &+ \left. \ \left(\frac{1}{2}-\beta_0\right)\int_0^\psi (\psi-E)^{-1/2-\beta_0}\,F_0(E)\,\txd E
  \right].
  \label{dfpsi}
\end{align}
Let us examine the first term inside the brackets: if
\begin{equation}
  \lim_{E\rightarrow\psi}(\psi-E)^{1/2-\beta_0}\,F_0(E) > 0,
\end{equation}
then
\begin{equation}
  \lim_{E\rightarrow\psi}(\psi-E)^{-1/2-\beta_0}\,F_0(E) \sim
  \lim_{E\rightarrow\psi}(\psi-E)^{-a} \text{\quad with $a\geq 1$},
\end{equation}
so that
\begin{equation}
  \int_0^\psi (\psi-E)^{-1/2-\beta_0}\,F_0(E)\,\txd E = +\infty.
\end{equation}
In other words, if the limit is nonzero, then the integral in the
second term becomes infinite. The limit can therefore be omitted, so that the 
equation is simplified to
 \begin{equation}
 \frac{\txd f}{\txd \psi}(\psi) = (2\pi)^{3/2}2^{-\beta_0}M\frac{\Gamma(1-\beta_0)}{\Gamma(1/2-\beta_0)}
 \int_0^\psi \frac{F_0(E)}{(\psi-E)^{1/2+\beta_0}}\,\txd E \geqslant 0,
\label{betalt1/2}
 \end{equation}\revision{
 and recalling Eq.~(\ref{gdsai}), the GDSAI is proven.
 The above relation can be generalized further: if $n = \lfloor 3/2-\beta_0
 \rfloor$ and $\alpha = 3/2 - \beta_0 -n$ are the integer floor and
 fractional part of $3/2-\beta_0$, then
 \begin{equation}
 \frac{\txd^k f}{\txd \psi^k}(\psi) = (2\pi)^{3/2}2^{-\beta_0}M\frac{\Gamma(1-\beta_0)}{\Gamma(3/2-\beta_0-k)}
 \int_0^\psi (\psi-E)^{1/2-\beta_0-k}\,F_0(E) \geqslant 0,\qquad 0 \leqslant k \leqslant n,
 \label{betalt1/2gen}
 \end{equation}
so the inequalities
  \begin{equation}
    \frac{\txd^k f}{\txd \psi^k}(\psi) \geqslant 0,
    \quad\qquad \forall\ 0 \leqslant \psi \leqslant \psi_0,\quad 0 \leqslant k \leqslant n,
  \end{equation}
  are necessary conditions to obtain a separable system with a non-negative
  DF. This extends the results obtained by \cite{2010MNRAS.401.1091C} for
  multi-component Cuddeford models.}

\subsection{Proof for $\beta_0 = 1/2$}

When $\beta_0 = 1/2$, Eq.~(\ref{fpsi}) reduces to
\begin{equation}
  f(\psi) = 2\pi^2 M \int_0^\psi \,F_0(E)\,\txd E.
\end{equation}
The derivative is then simply
\begin{equation}
\frac{\txd f}{\txd \psi}(\psi) = 2\pi^2 M\,F_0(\psi) \geqslant 0,
\label{betaeq1/2}
\end{equation}
so evidently, the GDSAI is again a necessary condition for a physical
dynamical model.

\subsection{Counter-examples for $\beta_0 > 1/2$}
\label{counter.sec}
The proof is not applicable to systems with $\beta_0 > 1/2$. Indeed,
the derivative has the same form as Eq.~(\ref{dfpsi}), but now the
two terms inside the brackets are respectively $+\infty$ and $-\infty$
when $F_0(E)>0$, so their sum is undetermined. 
\revision{However, we can rewrite
Eq.~(\ref{fpsi}) using integration by parts as
\begin{equation}
  f(\psi) = (2\pi)^{3/2}2^{-\beta_0}M
    \frac{\Gamma(1-\beta_0)}{\Gamma(5/2-\beta_0)}
  \left[
    \psi^{3/2-\beta_0}\,F_0(0)
    + \int_0^\psi (\psi-E)^{3/2-\beta_0}\,F_0'(E)\,\txd E
  \right],
\end{equation}
where $F_0'(E)$ denotes the derivative of $F_0(E)$. After
differentiation, we then obtain
 \begin{equation}
 \frac{\txd f}{\txd \psi}(\psi) 
 = (2\pi)^{3/2}2^{-\beta_0}M
 \frac{\Gamma(1-\beta_0)}{\Gamma(3/2-\beta_0)}
 \left[
   \psi^{1/2-\beta_0}\,F_0(0)
    + \int_0^\psi (\psi-E)^{1/2-\beta_0}\,F_0'(E)\,\txd E
  \right].
\label{betagt1/2}
\end{equation}
Thus, separable systems with a monotonically increasing $F_0(E)$
(i.e.\ $F_0'(E)\geq 0\quad \forall E$), satisfy the GDSAI. Again, this is
an extension of the results for generalized Cuddeford systems found by
\cite{2010MNRAS.408.1070C}.} 

Yet, the GDSAI is no longer a necessary condition for a physical
model, which raises the question whether systems can be found for
which the global inequality does not hold. To this aim, we consider
the potential-density pair
\begin{align}
  \psi(r)
  &=
  \frac{GM_{\text{tot}}}{\left(1+\sqrt{r}\right)^2}, \label{halo}
  \\
  \rho(r)
  &=
  \frac{3M}{8\pi}\,
  \frac{1}{r^{3/2}\,\left(1+\sqrt{r}\right)^4}, \label{rhomoore}
\end{align}
with corresponding density slope
\begin{equation}
\gamma(r) = \frac{3/2 + 7/2\sqrt{r}}{1 + \sqrt{r}},
\end{equation}
which is part of the family of Veltmann models or $\alpha$-models
(\cite{1979AZh....56..976V}; \cite{1996MNRAS.278..488Z}), and was
discussed by \cite{1998ApJ...499L...5M}. If $M_{\text{tot}}=M$, then
the system is also self-consistent. For this pair, we construct
physical DFs that generate four-parameter anisotropy profiles of the form
\begin{equation}
  \beta(r)
  =
  \frac{\beta_0+\beta_\infty(r/r_\txa)^{2\delta}}{1+(r/r_\txa)^{2\delta}},
\label{betagen}
\end{equation}
with $0<\delta\leqslant 1$, so that
\begin{equation}
  \tilde\rho(\psi,r)
  =
  f(\psi)
  \left(\frac{r}{r_\txa}\right)^{-2\beta_0}
  \left(1+\frac{r^{2\delta}}{r_\txa^{2\delta}}\right)^{\beta_\delta},
\label{rhofox}
\end{equation}
with
\begin{equation}
  \beta_\delta
  =
  \frac{\beta_0-\beta_\infty}{\delta}.
\end{equation}
Again, our systems have separable augmented densities. For every
anisotropy profile, the function $f(\psi)$ follows from $\rho(r) =
\tilde\rho(\psi(r),r)$, and the DF can be found by inverting
Eq.~(\ref{rhodf}). Instead of performing these calculations directly,
we adopt the technique used in \citetalias{2009ApJ...690.1280V}: we first
generate a family of components of the form
\begin{equation}
  \tilde\rho_i(\psi,r)
  =
  \rho_{0i}
  \left(\frac{\psi}{\psi_0}\right)^{p_i}
  \left(1-\frac{\psi^{s_i}}{\psi_0^{s_i}}\right)^{q_i}
  \left(\frac{r}{r_\txa}\right)^{-2\beta_0}
  \left(1+\frac{r^{2\delta}}{r_\txa^{2\delta}}\right)^{\beta_\delta},
\label{rhofam}
\end{equation}
where $p_i$, $q_i$ and $s_i$ are three parameters, and $\rho_{0i}$ are
normalization constants. The corresponding DFs are (see
\citetalias{2007A&A...471..419B}; \citetalias{2009ApJ...690.1280V})
\begin{align}
  F_i(E,L)
  &=
  \frac{\rho_{0i}}{M(2\pi\,\psi_0)^{3/2}}\,
  \sum_{j=0}^\infty
  (-1)^j\,
  \binom{q_i}{j}\,
  \left(\frac{E}{\psi_0}\right)^{p_i+js_i-3/2}
  \nonumber \\
  &\times
  \sum_{k=0}^\infty
  \binom{{\beta_\delta}}{k}\,
  \dfrac{\Gamma(1+p_i+js_i)}
  {\Gamma\left(1-\beta_0+k\delta\right)\Gamma\left(p_i+js_i-\frac{1}{2}+\beta_0-k\delta\right)}
  \left(\dfrac{L^2}{2r_\txa^2E}\right)^{-\beta_0+k\delta}
\label{dfbase1}
\end{align}
for $L^2<2r_\txa^2E$, and
\begin{align}
  F_i(E,L)
  &=
  \frac{\rho_{0i}}{M(2\pi\,\psi_0)^{3/2}}\,
  \sum_{j=0}^\infty
  (-1)^j\,
  \binom{q_i}{j}\,
  \left(\frac{E}{\psi_0}\right)^{p_i+js_i-3/2}
  \nonumber \\
  &\times
  \sum_{k=0}^\infty
  \binom{{\beta_\delta}}{k}\, 
  \dfrac{\Gamma(1+p_i+js_i)}{
    \Gamma\left(1-\beta_\infty-k\delta\right)
    \Gamma\left(p_i+js_i-\frac{1}{2}+\beta_\infty+k\delta\right) }
  \left(\dfrac{L^2}{2r_\txa^2E}\right)^{-\beta_\infty-k\delta}
\label{dfbase2}
\end{align}
for $L^2>2r_\txa^2E$. With different values of the
parameters, a library of base functions is thus created, from which a
linear combination is built that fits the given density $\rho(r)$ at
various radii. This is achieved by minimizing the quantity
\begin{equation}
  \chi_{N}^2 
  = 
  \frac{1}{N_{\text{data}}}\sum_{m=1}^{N_{\text{data}}}\frac{1}{\rho(r_m)}
  \left(\rho(r_m) -
  \sum_{i=1}^{N} a_i\rho_{i}(r_m)\right)^2,
\end{equation}
using a quadratic programming algorithm
\citep{1989ApJ...343..113D}. The details of this procedure can be
found in \citetalias{2009ApJ...690.1280V}. In particular, we created
models with $N=12$ components, fitting 25 density data points extracted
from Eq.~(\ref{rhomoore}).

\begin{figure*}[t]
\centering
\includegraphics{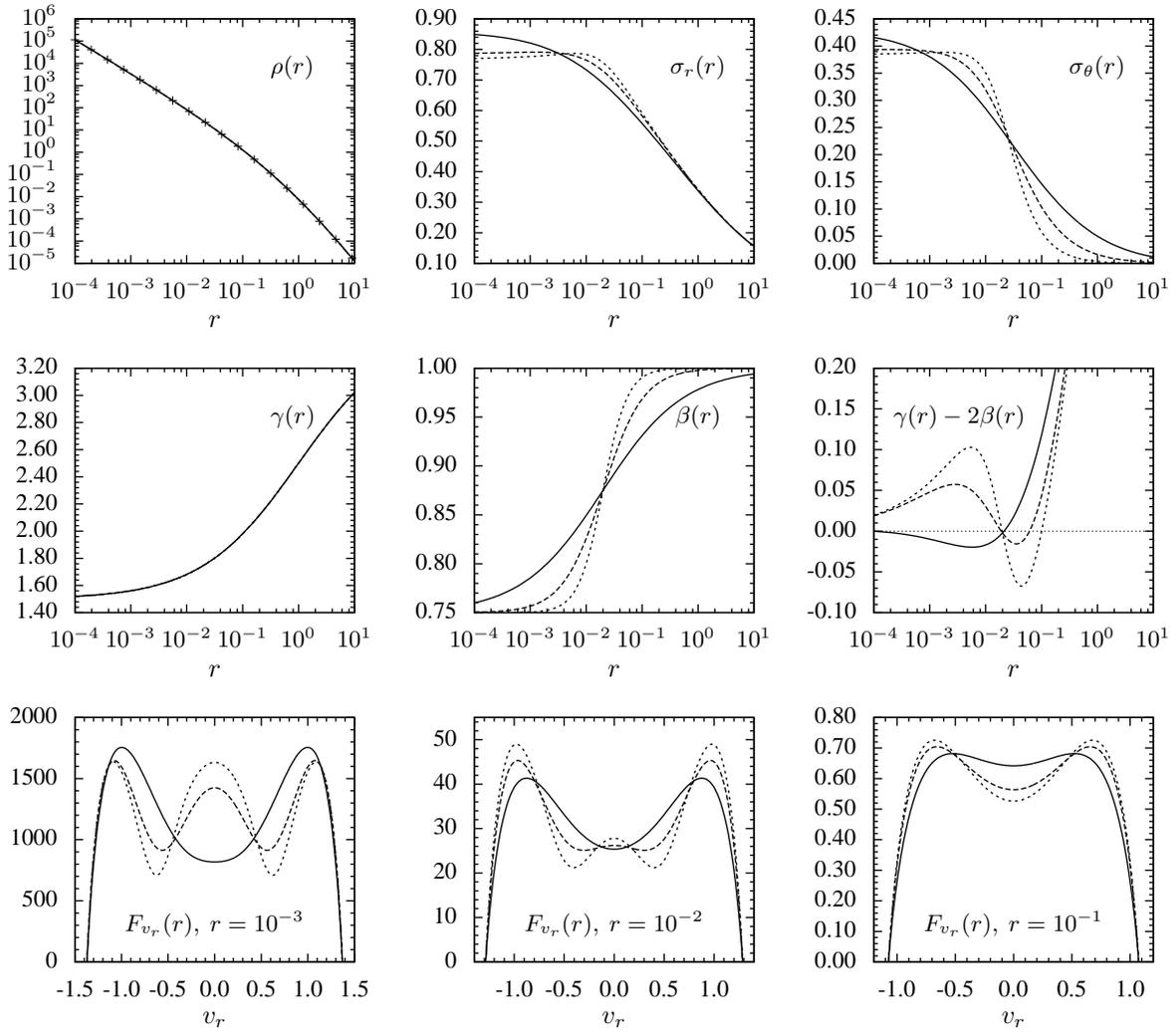}
\caption{Three models for which the GDSAI does not hold: $\delta=0.3$ (solid line), 
$\delta=0.6$ (dashed line), and $\delta=1.0$ (dotted line). In the first panel, the
density data points are also displayed.}
\label{figure1.sec}
\end{figure*}

With this technique, we obtain several dynamical models with
non-negative DFs that violate the GDSAI; three of them are shown in
Fig.~\ref{figure1.sec}. All three share the anisotropy parameters
$\beta_0=0.75$, $\beta_\infty=1$ and $r_\txa=0.02$, but have different
values for $\delta$: 0.3, 0.6 and 1.0 respectively; note that the
latter is a Cuddeford-type model. For the model with $\delta=0.3$, we
find that $\gamma(r)<2\beta(r)$ for radii in the interval $]0,0.021]$,
with a minimum around $r=0.0057$ (note that the center is a local
maximum, for which $\gamma_0=2\beta_0$). In the model with
$\delta=0.6$, the $\gamma-\beta$ relation reaches a local maximum
around $r=0.0028$, and the GDSAI does not hold in the interval
$[0.019,0.061]$, with a minimum around $r=0.036$. Finally the largest
$\gamma-\beta$ fluctuations occur in the Cuddeford model ($\delta=1$),
with a local maximum around $r=0.0054$, and a GDSAI violation within
$[0.019,0.100]$, with a minimum for $r=0.044$.

Evidently, we require rather extreme parameter
values to obtain these (modest) violations, while maintaining
non-negative DFs. The central anisotropy $\beta_0$ has to be high, and
the profile $\beta(r)$ has to increase very rapidly. It is therefore
safe to assume that the self-consistent variants of these models are
dynamically unstable. This can be seen from the standard criterion for
radial-orbit instability: $2T_r/T_T=2\langle v_r\rangle/\langle
v_T\rangle =5.45,$ 8.26 and 10.42 for the three models, which is much
higher than the $\simeq 2$ threshold for similar models (see
\citet{1999PASP..111..129M} for an overview). Further evidence 
of dynamical instability is given by the radial
velocity distributions
\begin{equation}
F_{v_r}(r) = 2\pi M \int_0^{\sqrt{2\psi(r)-v_r^2}} F(E,L)\,v_T\,\txd v_T.
\end{equation}
As shown in the bottom row of Fig.~\ref{figure1.sec}, these profiles
have two or three peaks at small radii. These are indications of
H\'enon instabilities (see \citet{1999PASP..111..129M};
\citet{1986ApJ...300..112B}). In theory, if the systems are instead
not self-consistent but embedded in a massive dark matter halo, they
might withstand these instabilities; however, one can safely argue that
such equilibrium systems are too extreme to arise in structure
formation.

\subsection{The inverse relation}
\revision{
Finally, we remark that the function $F_0(E)$ can be derived from $f(\psi)$ by means of
  an Abel-related inversion
  \citep{1991MNRAS.253..414C,2006AJ....131..782A}, which holds for all values of
  $\beta_0<1$,
 \begin{equation}
   F_0(E) = \frac{2^{\beta_0}}{(2\pi)^{3/2}M\Gamma(1-\alpha)\Gamma(1-\beta_0)}
   \left( \int_0^E \frac{\txd^{n+1}f}{\txd\psi^{n+1}}\frac{\txd\psi}{(E-\psi)^\alpha}
     + \frac{1}{E^\alpha}\frac{\txd^{n}f}{\txd\psi^{n}}(0) \right),
 \end{equation}
where again $n = \lfloor 3/2-\beta_0
 \rfloor$ and $\alpha = 3/2 - \beta_0 -n$ are the integer floor and
 fractional part of $3/2-\beta_0$.
Thus the additional condition
  \begin{equation}
    \frac{\txd^{n+1} f}{\txd \psi^{n+1}}(\psi) \geqslant 0,   
    \qquad \forall\ 0 \leqslant \psi \leqslant \psi_0,
    \label{sufcond}
  \end{equation}
  is sufficient to obtain a non-negative $F_0(E)$. As
  \cite{2010MNRAS.401.1091C} showed, this also implies that the entire
  DF $F(E,L)$ is non-negative in the case of generalized Cuddeford
  systems. However, it is not a priori clear whether this property is
  true for all separable systems, since the behavior of $F_1(E,L)$
  might still lead to negative values of the DF. Further study is
  therefore needed to determine if Eq.~(\ref{sufcond}) is a sufficient 
  condition for the existence of a physical separable model.}

\section{Discussion}
\label{conclusions.sec}
In the previous section, we presented a full analysis of the GDSAI for
spherical dynamical systems with a separable augmented density. As our
proof shows, the GDSAI hold if the central velocity anisotropy
$\beta_0\leq 1/2$. We further demonstrated that systems with $\beta_0
> 1/2$, the GDSAI can be broken, as shown by three counter-examples,
although these systems are not physically realistic.

Eqs.~(\ref{betalt1/2}) and (\ref{betaeq1/2}), combined with
Eq.~(\ref{gdsai}), also reveal a remarkable property of separable
systems: the GDSAI is purely determined by $\beta_0$ and $F_0(E)$. The
function $F_0(E)$ can be interpreted in various ways: it can be
thought of as the phase-space distribution of particles at purely
radial orbits, as the phase-space distribution of particles at the
center, or as the energy distribution of the constant-anisotropy
component of the DF. \revision{As a surprising consequence, if a
  separable system has a given potential $\psi(r)$ and density
  $\rho(r)$, then knowledge of $F_0(E)$ alone is sufficient to
  construct the complete DF of the system. Indeed, we showed that
  $F_0(E)$ is equivalent with $f(\psi)$, and in combination with
  $\rho(r)$, the function $g(r)=\rho(r)/f(\psi(r))$ can also be
  derived, determining the augmented density $f(\psi)\,g(r)$ and thus
  $F(E,L)$.}

The next logical step will be to investigate the GDSAI for general,
non-separable spherical models. One possible approach would be to
consider a spherical systems as a linear combination of separable
systems. In fact, an analytic $\tilde\rho(\psi,r)$ or $F(E,L)$ can be
written as a double sum of power-law functions, by means of a
two-dimensional Laurent series expansion. An alternative approach
would be to ask the following question: given a spherical dynamical
system with a given $\psi(r)$ and a DF that generates $\rho(r)$ and
$\beta(r)$, does there always exist a separable model with a
non-negative DF that generates the same density and anisotropy?  As we
mentioned in Section~\ref{augdens.sec}, the function $g(r)$ follows
directly from $\beta(r)$, and in turn this determines $f(\psi)$ from
$g(r)$ and $\rho(r)$. However, there is no a priori reason that the
corresponding DF is also non-negative. If it is, the same GDSAI
analysis applies as presented in this paper. We currently do not know
of any $(\rho(r),\beta(r))$ pair that can only be generated by
non-separable models, but more study is required to resolve these
questions. We think our analysis of separable systems can be a useful
stepping stone for further investigations of the GDSAI for spherical
dynamical systems.

\acknowledgments{
\revision{The authors wish to thank the referee Luca Ciotti for the
generous comments and helpful suggestions that improved our paper.
}
}
 
\end{document}